\documentclass{acm_proc_article-sp}

\usepackage{graphicx}
\usepackage{mathtools}

\begin{document}

\title{The challenge of decentralized marketplaces}
\subtitle{Bas van IJzendoorn, 4024850\\ Delft University of Technology, Distributed Systems group }
\author{Responsible faculty member: Johan Pouwelse}

\maketitle
\begin{abstract}
Online trust systems are playing an important role in to-days world and face various challenges in building them. Billions of dollars of products and services are traded through electronic commerce, files are shared among large peer-to-peer networks and smart contracts can potentially replace paper contracts with digital contracts. These systems rely on trust mechanisms in peer-to-peer networks like reputation systems or a trustless public ledger. In most cases, reputation systems are build to determine the trustworthiness of users and to provide incentives for users to make a fair contribution to the peer-to-peer network. The main challenges are how to set up a good trust system, how to deal with security issues and how to deal with strategic users trying to cheat on the system. The Sybil attack, the most important attack on reputation systems is discussed. At last match making in two sided markets and the strategy proofness of these markets are discussed.
 \nocite{*}
\end{abstract}

\section{Introduction to Online Trust Systems}
We focus on enabling trust and transactions between peers in a decentralized market. A decentralized market is a market where traders directly meet each other without an intermediate party like in a centralized market. Removing these intermediaries reduces cost. Williamson (1993)\cite{Williamson} distinguishes six types of trust contexts that are important for economic activity: societal trust, political trust, regulatory trust, professional trust, network trust and trust in the corporates themselves. As these contexts are outside of the digital world it is hard for a computer to calculate whether another party is trustworthy or not. The lack of social relations with other traders in markets on the Internet can create problems in communication and trust (Furlong, D., 1996)\cite{Furlong}. For instance, in grain trading markets, sharing all information gives opportunities for actors to use this information and exploit the sharing trader. A buyer in grain trading markets might be reluctant in sharing how much grain it wants to buy because this gives valuable information about the trading position of the buyer. When other actors know the trading position of the buyer they can play economic games like only selling grain to this buyer for a higher price.

The feeling of trust a trader has with another trader is not purely rational. In neo-classical economics traders come together for an instant to exchange goods to maximize their utility in the perfect market. The traders maximize their utility by rational reason. In later research in economics and politics, researchers came to the consensus that actors had a bounded rationality and abandoned the idea of decisions based on reason. Actors cannot oversee the choice of possibilities and options to rationally maximize utility (Simon, H.A., 1972)\cite{BoundedRationality}. Economists began to think of about other models that explain the economic behaviour of actors. Therefore social relations and trust were introduced as concepts in economic decision making research (Furlong, D., 1996)\cite{Furlong}.

There are examples of markets where a lot of volume is traded via the Internet and where the trust relation appears not to be a problem. For instance, the volume traded in decentralized markets like used cars, used books or used furniture has increased dramatically in the recent years. In the period 1997-2007 the volume increase of used cars traded increased by 7.2 procent in California. This implies a welfare gain of $\$$43 million per year relative to 1997 in California alone. Internet allows targeted search in niche markets like used automobiles. Information about the product is shared free of charge to all potential buyers. The geographical reach that Internet provides to potential buyers is the main reason for the success of such markets. Also, the transaction cost of a trade decreases in Internet based decentralized markets compared to centralized markets. Transactions costs are the costs of finding the right information, bargaining costs and contract costs. (Rapson and Schiraldi, 2013)\cite{Rapson}. What kind of information and how information should be presented to users depend on the structure of the market. In the survey multiple types of markets are researched, each with their own characteristics. \cite{lucking2001business} \cite{wigand1997electronic}

\section{Operational systems}
Online markets started 22 years ago. Numerous examples exist which reached large-scale usage. Such operational systems provide solid proof for our design space.

\subsection{Decentralized markets}
Pierre Omidyar first started the peer-to-peer decentralized market eBay when he put a broken laser pointer for sale and sold it for \$14.83. eBay allows buyers to buy all sorts of goods online from sellers throughout the world. Other markets like computer programming (Freelancer, oDesk), consumer loans (Prosper, Lending Club), crafts (Etsy), start-up financing (Kickstarter), accommodation (Airbnb), baby-sitting (Care.com), currency exchange (Transferwise, CurrencyFair) and on demand rides (Uber, Lyft, Blabla Car). They match buyers and sellers or implement an auction-based pricing. In 1995 people still were reluctant to send money across the country to an anonymous seller for a product on eBay. \cite{einav2016peer} \cite{oskam2016airbnb} \cite{lehr2015analysis}

Trust is vital for commerce. The electronic commerce acceptance model explains when users have the intent to go to an electronic shop instead of a real physical one. It turns out that not only the information quality, service quality and system quality have its effect on the perceived usefulness of the system, but also the perceived trust of an e-commerce website plays a vital role in the perceived usefulness of the system and the altitude of the system. Successful e-commerce websites ensure a low level of consumer risk perception and a high level of consumer trust perception. \cite{corbitt2003trust} In some markets trust is created by inspection or by external regulations. For instance, sellers will post pictures of the product on eBay and eBay compensates the buyer if the seller does not deliver as advertised. The most used trust mechanism is reputation or feedback systems: users give feedback after the delivery of a product or service. In eBay most bad actors and highly fraudulent behavior is filtered out with the reputation system. \cite{einav2016peer} \cite{ratnasingham1998importance} \cite{gefen2000commerce}

\subsection{Decentralized exchanges}
Some markets have an exchange where commodities, financial products, currencies and futures are traded. Sellers provide ask prices and buyers provide bid prices. Exchanges require a centralized component through which the products are traded. The first attempt at a decentralized exchange market is Tsukiji, by The,M. and Reinbergen, H. (2013). It is a simple implementation where decentralized nodes act as traders. The traders can place bid and ask offers and respond to an offer such that a trade can be established. The discovery of peers is also implemented but there is no real money traded and there also isn't a working user interface. \cite{Tsukiji}

An improvement on the design of Tsukiji is the Decentral exchange market design by Olsthoorn, M.J.G. and Winter, J. (2016). Instead of peer discovery bid and ask prices together with quantities are distributed across the network with ticks when a peer bids or asks a certain quantity. Secondly, there is a simple matching engine implemented that matches bid and ask  quantity amounts with the highest and lowest prices.  Then when a match is made real money is traded. MultiChain coins of Tribler peers are traded against BitCoins in a single transaction where both wallets of both traders are updated. The design is successfully implemented in Tribler, constructed with Dispersy and tested. \cite{OlsthoornMarket}

\begin{figure}
  \includegraphics[width=\linewidth]{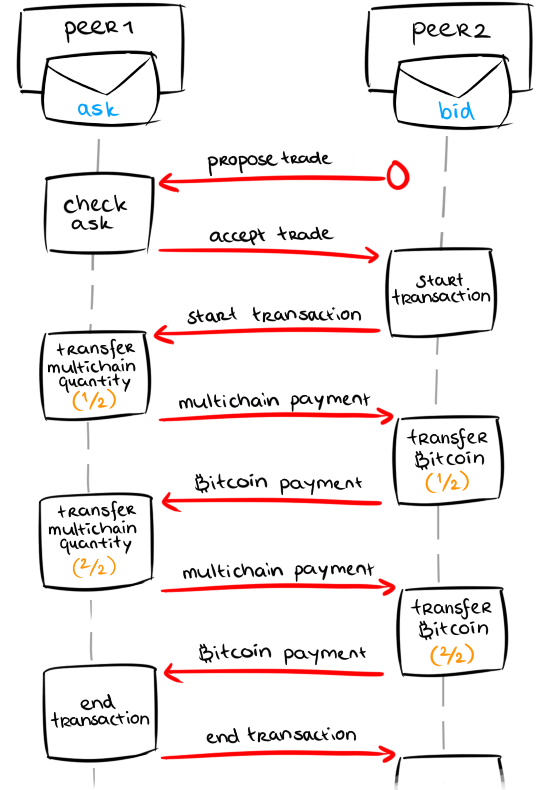}
  \caption{Decentralized exchange system by Olsthoorn and Winter (2016) \cite{OlsthoornMarket}}
\end{figure}

\begin{figure}
  \includegraphics[width=\linewidth]{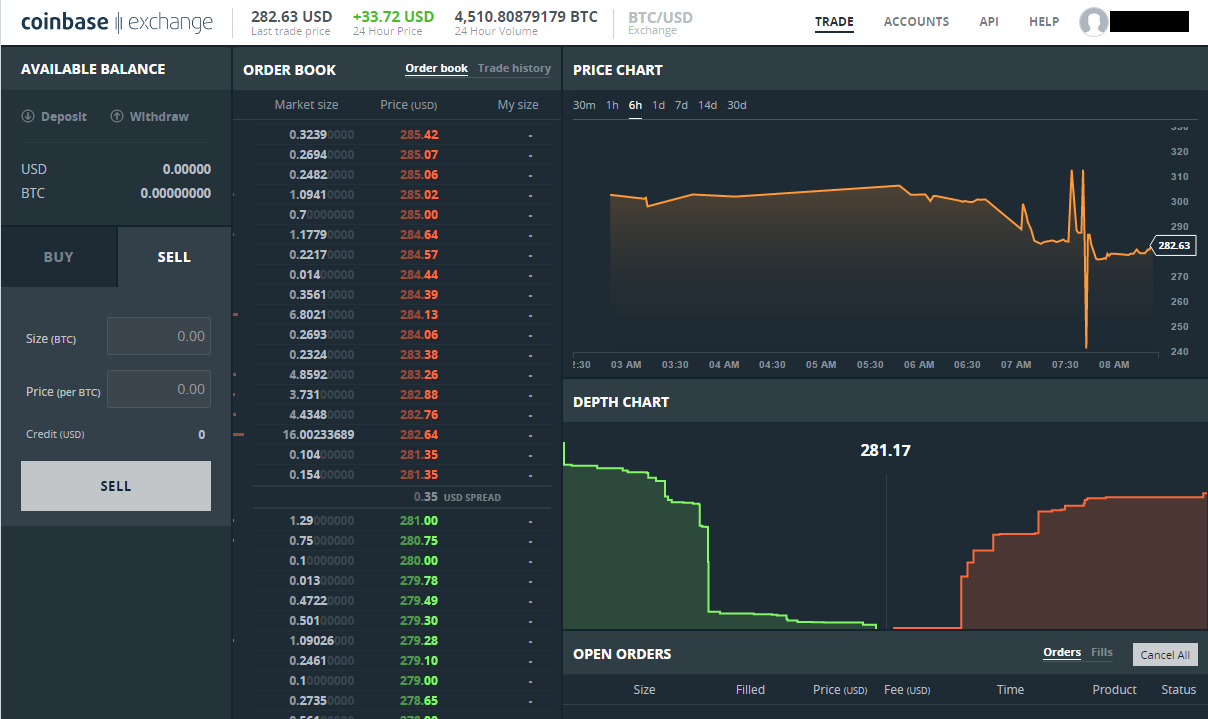}
  \caption{Bitcoin / Dollar exchange example download from www.coinbase.com}
\end{figure}

\subsection{Anonymous decentralized markets}
Anonymous markets make perfect illegal decentralized markets. On the Silk Road there is a mechanism in place where only 2.2\% of the transactions are fraudulent. There appears to be honesty among illegal drug dealers. The Silk Road marketplace is an independent marketplace where buyers and seller conduct in electronic commerce transactions. Using TOR technology the Silk Road also provides anonymity for its users. Items are payed with bitcoins. Most items being sold on the Silk Road are illegal narcotics such as Weed, Drugs, Cannabis, Cocaine and Pills where most of the items come from the U.S.A. (43,86\%), U.K. (10.14\%) and the Netherlands (6,51\%). The items are delivered worldwide. Interestingly the transaction volume stays about the same while the bitcoin price changed. The number of sellers doubled almost in 6 months time from February to august 2012. Most of the new sellers leave the site fairly quickly. Only about 4\% of the sellers have been on the site for the entire duration of the measurements in 2012. Because of the illegal items that are being sold on the Silk Road some of the the Silk Road got eventually taken down by law enforcement. \cite{Christin:2013:TSR:2488388.2488408} \cite{soska2015measuring}

\begin{figure}
  \includegraphics[width=\linewidth]{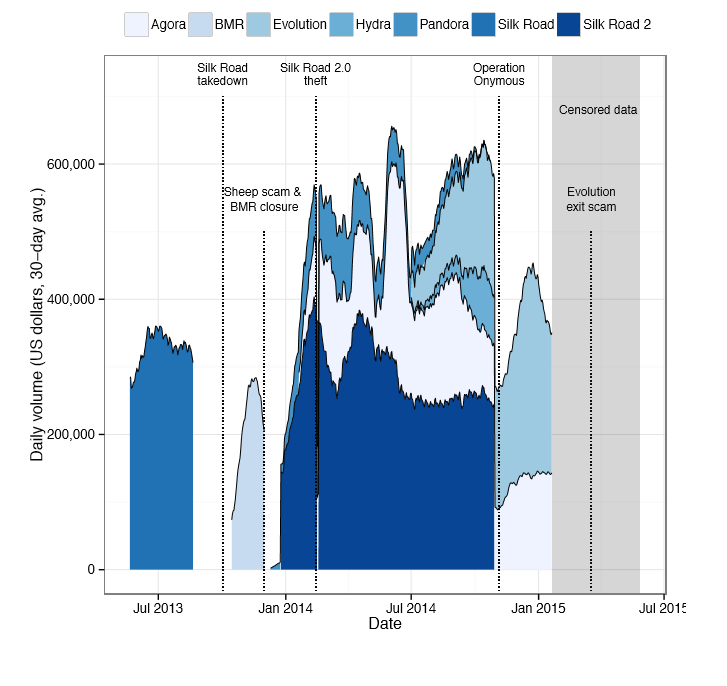}
  \caption{Sales volume in the entire anonymous decentralized market ecosystem \cite{soska2015measuring}}
\end{figure}

Between september 2013 and january 2015 new versions of the Silk Road emerged like Agora, BMR, Evolution, Hydra, Pandora and the Silk Road 2. Some were also taken down by law enforcement after some time. The total volume through these illegal markets remains roughly the same despite the interference of law enforcement. Also the fractions of sales per item category also remained the same throughout the period. Sellers simply made new aliases at different marketplaces. The total number of vendors increased when marketplaces were introduced and decreased again when marketplaces were taken down. It appears that the sales of illegal items is independent of what marketplace is used and interference by law enforcement. \cite{soska2015measuring}

\subsection{Trust in P2P file sharing}
BitTorrent is a P2P file sharing system. It has a mechanism in place to measure the trustworthiness of users and to prevent free-riding behavior where users only download files from the system that is inspired on Tit-for-Tat. A peer in BitTorrent prefers to upload more data to another peer it has downloaded from. Uploading to another peer thus increases the trustworthiness of that peer. In Kazaa a more complicated mechanism is in place where some peers are elected super nodes and peers receive peer-points for uploads. Super nodes get more responses of peers who spend their peer-points to gain a higher download speed.\cite{tamilmani2006studying} \cite{cohen2003incentives} \cite{marti2006taxonomy}

At Delft University of Technology a dedicated blockchain is operating for creating trust. Two trust schemes have been tested in Tribler: Bartercast and Multichain. Tribler is a P2P file sharing system used for research. In BarterCast, a peer collects upload and download speeds of other peers by requesting this information from peers. The information received is then forwarded to ten other peers. By this way upload/download ratio information is shared among all peers and a map of peers with their ratios can be created. A bloom filter algorithm is in place that deletes duplicate information. Each peer can calculate the reputation values of other peers with the max-flow algorithm. In BarterCast there is no global reputation value calculated by an authority. Every peer maintains its own list of reputations of other peers. It is assumed that no cheating is done upon the sharing of information. Truth-telling of nodes is assumed and the trustworthiness of nodes is assumed to be high. It is easy to attack BarterCast by simply stating a high upload amount to other users. Sybils can verify this high upload amount to help fool honest nodes (Meulpolder,M. et al, 2009). \cite{meulpolder2009bartercast} \cite{meulpolder2008bartercast}

\begin{figure}
  \includegraphics[width=\linewidth]{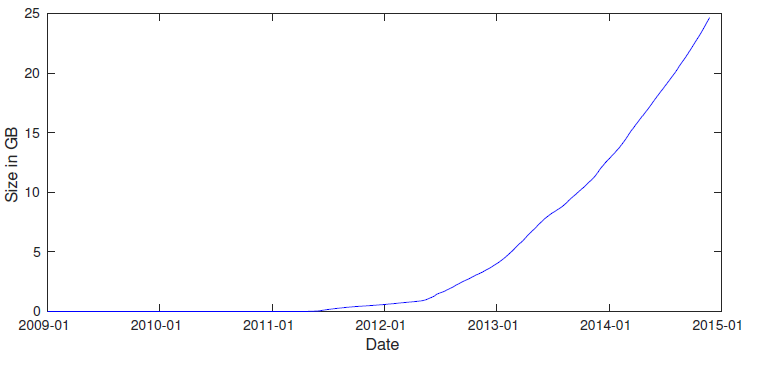}
  \caption{Increasing size of BlockChain \cite{MultiChain}}
\end{figure}

MultiChain is an improvement on BarterCast. A payment system is introduced that replaces BarterCast completely. The payment protocol and datastructure is inspired by the BlockChain payment technology. In both MultiChain and in BlockChain a chain of blocks with transactions is maintained to prevent the double-spending of coins. The difference between BlockChain and MultiChain is that MultiChain blocks are distributed among the two peers of the interaction instead of one single BlockChain. The benefit of this is that the ChainSize is kept small in MultiChain. In BlockChain the size of the chain is ever increasing and becomes a inoperable after some time. But MultiChain also introduces some of its own problems. When a node fails MultiChain cannot check anymore for double spending. The coin could be traded by the failing node and the failing node is the only node that knows where the coin goes next. Also transactions cannot be performed fast after each other. A transaction has to be processed completely in a block before a new transaction can be made. This gives scalability problems (Norberhuis, S., 2015). \cite{MultiChain} \cite{nakamoto2008bitcoin} \cite{blundell2014bitcoin}

Tribler also uses decentralized credit mining to gain trust in other P2P file sharing networks. The system aims to earn trustworthiness of peers in other swarms. In the paper by Capota et al (2015)\cite{DecentralizedCreditMining} this is described as earning credit in other swarms on behalf of the user. The system is part of the Tribler P2P client and is implemented for every peer and therefore completely decentralized. The system selects swarms on its upload potential and start to upload data to these swarms. In this way the peer gains trust in that swarm. Information is frequently updated to maximise upload to swarms and there are also spam detection and duplicate content detection to further enhance the upload process. The system is also tested to show that trust is gained in other swarms with the system. The underlying mechanism to gain trust in the paper is simple. The peers simply behave cooperatively by uploading data to proof that they are not free riders and thus to proof their trustworthiness.

\subsection{Ethereum: Smart contracts}
With smart contracts people are able to execute trades through Trustless public ledgers (TPLs). TPLs allow a restructuring of power relations between parties and intermediaries. TPLs enable parties to store digital assets on-line without the need of banking intermediary who charges a fee. In addition to that they also allow parties to transfer digital assets directly to each other on their own terms. The conditions of the terms can be programmed in a "smart contract": "an automated program that transfers digital assets with BlockChain technology upon certain triggering conditions". Smart contracts do not require an institution as an intermediary exchange. Smart contracts also solve the longstanding problem of e-commerce courts to refuse to protect consumer contract terms. With smart contracts consumers can express their own wishes for the contractual terms and negotiate with other parties on their own (Fairfield, 2014). \cite{fairfield2014smart}

A practical implementation of smart contracts is the Ethereum system. Money is traded with smart contracts using its own currency: "Ether". The underlying transactions of the smart contracts are done with BlockChain Technology. BlockChain does not only provide an infrastructure for digital payments, but also provides a distributed consensus for the rightness of the payments and prevents double spending attacks. Ethereum is a fully fledged Turing-complete programming language that can create a wide range of financial applications like smart contracts, digital currencies for exchange and also programmable decentralized autonomous organizations which are organizations where the money management of an organization is completely on the trustless public ledger(DAOs). \cite{buterin2013ethereum} \cite{wood2014ethereum} \cite{delmolino2015programmer}

Paper contracts could be replaced by Ethereum contracts. Paper contracts are an agreement between parties to do or not do something. For instance, a grain seller agrees on delivering an amount of grains to the Paranagua harbor in Brazil at a certain date and time. With Ethereum it is possible to handle the contract details on-line in the BlockChain. Egbertsen et al, (2016) \cite{egbertsen2016replacing} explains four fields of examples where Ethereum smart contracts could replace paper contracts. The first and probably the most widely used is a purchase agreement. In the current world money is put in Escrow at a third party upon a purchase. When both sides have fulfilled their parts, the purchase agreement is met and the money can be transferred. The third party can be ruled out with Smart Contracts. The other examples proposed by Egbertsen et al, 2016 are the certification of diploma's in the BlockChain, electronic voting and residential lease agreements. The Leonardo da Vinci Engineering School in Paris wants to issue their diploma's with smart contracts. Ukrainian officials are taking Ethereum serious as a voting system. The lease agreements could be stored on Ethereum to prevent altering of the contract. \cite{bogner2016decentralised} \cite{malkovsky2015concept} \cite{koulu2016blockchains}

\begin{figure}
  \includegraphics[width=\linewidth]{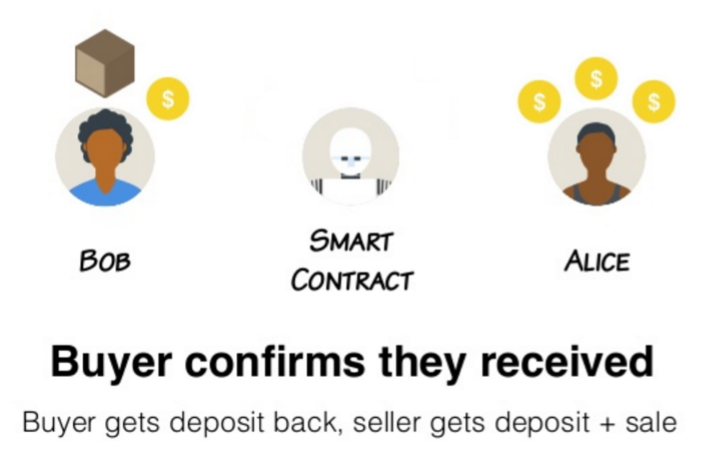}
  \caption{Purchase with Smart Contract \cite{egbertsen2016replacing}}
\end{figure}

\section{15 years of academic fantasy designs}
The academic world has been somewhat disconnected from operational systems.
Creating an experimental operational markets is difficult for academic institutions and merely obtaining detailed datasets is already problematic. 
This hampers the progress of the field in our opinion.

\subsection{Trust enforcements in P2P file sharing}
According to Moreton, T. (2003) \cite{TrustTokensStamps} the major problem in P2P systems is the mutual distrust between peers. There are many pseudonyms or Sybil nodes that take up resources without providing resources to the network. These Sybils are run by agents which have a bad trust relationship with the other agents of the network. The behaviour of these agents is in P2P filesharing also denoted as freeriding. The problem was first described by Wilcox O'Hearn (2002) \cite{OHearn} after his experiences with the deployment of the Mojo Nation file sharing system. O'Hearn also describes the mistrust among nodes as the biggest problem in Mojo Nation system. The motivation between nodes to cooperate was not there. Nodes did not upload data to the network which made data availability a problem. There were even attacks on the network by which users altered their clients to gain more advantage for himself. Users altered their clients to gain more trust in the system.

Tsuen-Wan et al (2003) \cite{wallach2003enforcing} proposed three solutions to the free-riding problem and to enforce sharing. Two of them are not suitable according to the authors. The third one introduces a method that involves the auditing of peer nodes. Each node maintains a usage file where it defines the amount of capacity it advertises and it also maintains the advertised capacities of all neighbours. A simple rule is added that says that a node can only download new data if its own advertised capacity is larger than the sum of the advertised capacity of all its neighbors. An auditing procedure is introduced that let nodes check on each other whether to tell whether they are trustworthy or not. The economics of the auditing model seems very unlikely to be successful. The required capacity needs to be very high to be able to download data. What's interesting about the paper is that the concept of an auditing procedure by other peers is introduced. By this way the network maintains its own reputation. 

Vishnumurthy, V. (2003) \cite{KARMA} introduces a design of a P2P file sharing system where a currency is introduced in where a single value called KARMA. The currency KARMA represents the amount of resources a peer has contributed and consumed in the network. This represents a users trustworthiness with regard to upload/download ratio within the system. The idea behind is that a user who has uploaded more is more likely to upload in the future and is therefore more trustworthy. This means other users can upload to this user and the user with high KARMA gets a higher download speed. The proposal of Vishnumurty is quite complex. There are groups of k nodes called bank-sets that keep track of the KARMA of each user. Mechanisms are in place to make the KARMA system work.  Distributed hash tables (DHT's) map nodes towards a bank set. When a node goes down, a new node becomes part of the bank set. It is impossible for nodes to adjust their KARMA level at will and KARMA can compensate bank nodes for participating in transactions with KARMA. Thus nodes who help in maintaining the system by banking get a small KARMA reward. There are also security mechanisms for replay attacks, malicious providers, malicious consumers, attacks against DHT routing, corrupt bank sets and denial of service attacks. However, KARMA does not protect against Sybil attacks. Protection against Sybil attacks will be discussed in a later section. 

\begin{figure}
  \includegraphics[width=\linewidth]{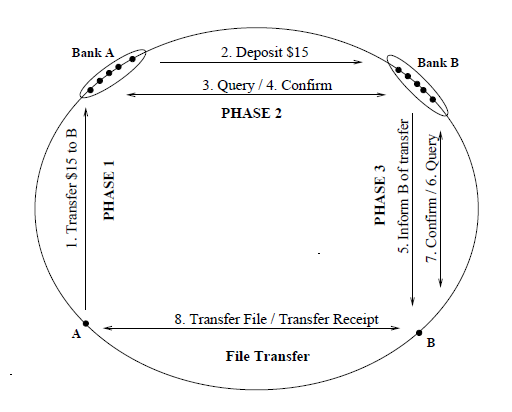}
  \caption{Karma-File exchange proposal from 2003\cite{KARMA}}
\end{figure}

A paper that tries to capture the essence of the combination of a reputation system and a payment protocol as with KARMA is the stamp trading model by Moreton et al (2003). \cite{TrustTokensStamps} Stamps are introduced that can be traded between nodes and can later redeemed at a node for service. In this payment protocol the stamps have a variable value and are traded based on this value. It is assumed there is a centralized exchange rate mechanism which can observe all interactions between node and thus provide perfect valuations to the stamps' value. This assumption has practical issues. In the first place it is hard to observe all interactions between nodes and secondly the centralized exchange rate node has to be trusted fully. If this central nodes gets compromised by an adversary, all interactions can be observed and the whole network is compromised. In the paper multiple price valuation methods are proposed with different properties. The schemes have to be both token-compatible and trust-compatible. A scheme is token-compatible if the total value of the stamps in the network is bounded. A scheme is trust-compatible if failure by a node to redeem a stamp never increases the total value of its stamps. In four of the proposed methods for pricing the system can be flooded with requests by nodes with a higher bandwidth to artificially obtain a higher trust. In the last method called Bounded Redemption Rate (BRR) the value of the stamp is chosen in such a way that flooding the network with stamps causes a node's total stamp value to approach zero value. In this way the BRR method becomes trust-compatible. It is also proven that BRR is also token-compatible. BRR can resist Sybil attacks because when a nodes becomes flooded with requests of pseudonyms, the total stamp value of a node approaches zero. However, stamp trading still has the following open problems: double spending, cryptographically signing stamps, audit trails of stamps, the token exchange problem which is now fixed with the central node assumption and limited knowledge on both the stamp-trading economies and attacks. Thus although stamp trading is resistant against some form of Sybil attacks it has many open problems which makes is impractical to implement in the real world.

\begin{figure}
  \includegraphics[width=\linewidth]{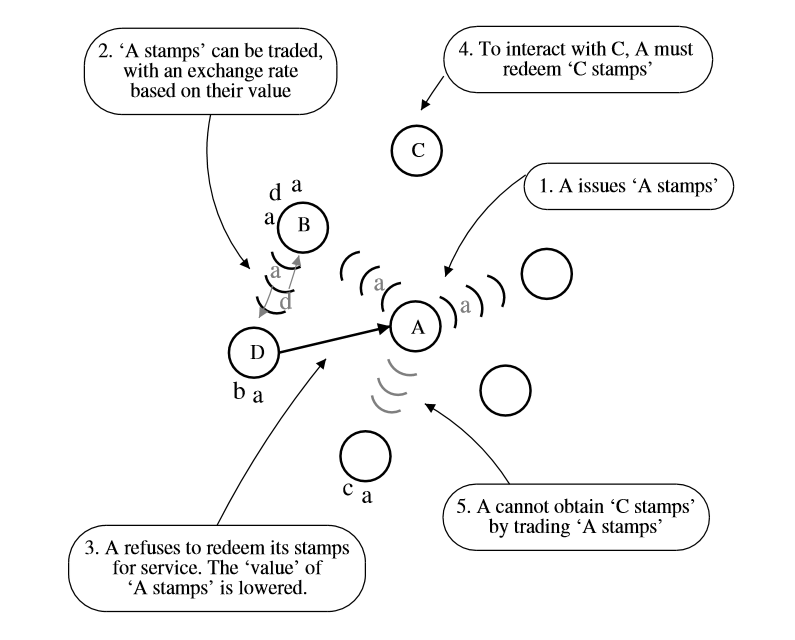}
  \caption{Stamp Trading Protocol from 2003 \cite{TrustTokensStamps}}
\end{figure}

PPay is a system introduced by Yang, B. et al (2003) \cite{ppay} that uses payment systems to fix the mutual distrust between peers. The solution Ppay is introduced in this paper and improves performance of micropayments while maintaining security. Unlike traditional transferable cash, coins in Ppay do not grow in size as they are transferred. A user purchases digital coins from a broker B. The user U is now the owner of the coin. U can assign the coin to another user V and V can do reassignment request to U. When V wants to reassign the coin to user X it has to go through the user of the coin U. Therefore U must always be online in order to reassign coins. To solve the problem of a potential crash of U there is a downtime protocol introduced that allows the holder of the coin to have the coin reassigned by the broker. In this case broker B will charge both U and V a percentage of the reassigned amount for this service. This charging gives incentives for nodes to remain online. The reassigning of the coin is computationally expensive. Ppay does not prevent coin fraud at the outset, but instead makes fraud unprofitable. Ppay ensures that any fraud can be detected and traced back to the misbehaving peer by means of an “audit trail” of the coin. The system can be attacked by replicating an assigned coin and spending it twice, wrongful denial and double spending. The broker will create the right punishments and will do risk management for the system. There 4 four issues and extensions described to solve certain problems. 1) Printing raw coins is expensive for the broker. This responsibility can be divided to users with limit certifications. 2) Layered coins: The coin transfer history is saved in layers at each coin. The reassignment adds a new layer to the coin. 3) Coin renewal: The audit trail is purged once in a while to limit the amount of state each peer should maintain. 4) Soft Credit Windows: Quick payments that go back and forth can be washed out. Payword hash chains are also a fast method. A quantitative analysis is performed that compares Ppay to RM. Ppay can significantly outperform existing schemes in terms of broker load, while maintaining a reasonable peer load. 

Ham, M. and Agha, G. (2005) \cite{ham2005ara} solve the trust problem in a similar way as with micro-payments with servers. The payment (credit) is made volatile and the approach does not rely on servers. It is assumed that a stricter system does not degrade its popularity because a system with free riders will eventually starve. Four types of cheating are targeted: Exaggerated credit by an individual peer, Conspiracy: a peer may evade detection using collaborators, Blame Transfer:  a cheater might blame an innocent peer to hide malicious peer misbehavior, Omitting Interested Peers: Omitting peers from malicious lists send to other peers. A credit system is introduced where credit is the uploaded bytes (contribution) minus the downloaded bytes (consumption). Two values LL and LLe are introduced as limits to the system as to when a peer should serve another peer. These limits solve the start-up deadlock and the starvation. 

Feldman, M.  et al (2004) \cite{FreeridingWhitewashing} made a mathematical model that studies the trust problem. The mathematical model has not been tested in the real world so nothing can be said about its validity. However, some useful observations can be extracted from the model. For instance, the behaviour of white-washers: users who leave the system and rejoin with new identities to avoid reputation penalties are added to the model. In the paper is not a new incentive scheme proposed. This is an example of a Sybil attack where pseudonyms leave the system and later rejoin to renew download speed. Sybil attacks are discussed later. 

70\% of Gnutella participants are free riders. e.a. Users that don't contribute to the system and are not trustworthy. With a lot of free riders the system does no longer provide utility to any of it's participants. The welfare of the system decreases to zero with a lot of free riders. A P2P contract is formulated by Ghosal et al, 2005 \cite{ghosal2005p2p} where a peer contributes an amount of resources $R$ to the system in exchange for some level of service $S$. There are other schemes to tackle the peer incentive problem. The first relies on altruism, this has worked for various systems such as Napster, Gnutella en Free net. However, they operate sub-optimally from the standpoint of maximizing the welfare of its participants. The second approach uses micropayments to provide economic incentives for resource contributions. In systems where the currency can be redeemed for goods outside of the system the problem of free riders does not exist because free riders have incentives to contribute. 

\subsection{Trust enforcements in TOR}
Free riding users with no trust are also a problem in anonymization networks that depend on a very small set of nodes that volunteer their bandwidth. In order to incentive bandwidth sharing Androulaki, E. (2008) \cite{PAR} proposes a design where payment systems are used that addresses problems such as the double spending problem with a hybrid payment scheme by combining features from the micropayment system and the e-cash scheme. The proposed scheme does not attempt to achieve absolute financial security but the authors are willing to accept small amounts of cheating. There are two types of coins in the proposal: S-coins and A-coins. S-coins are coins signed by relay nodes and are used to pay successor nodes in a circuit. A-coins are signed by the bank and bought by users to use the anonymization network.  S-coins can also be used to pay for using the anonymous network. This gives economic incentives for tor relays to forward traffic. 

\begin{figure}
  \includegraphics[width=\linewidth]{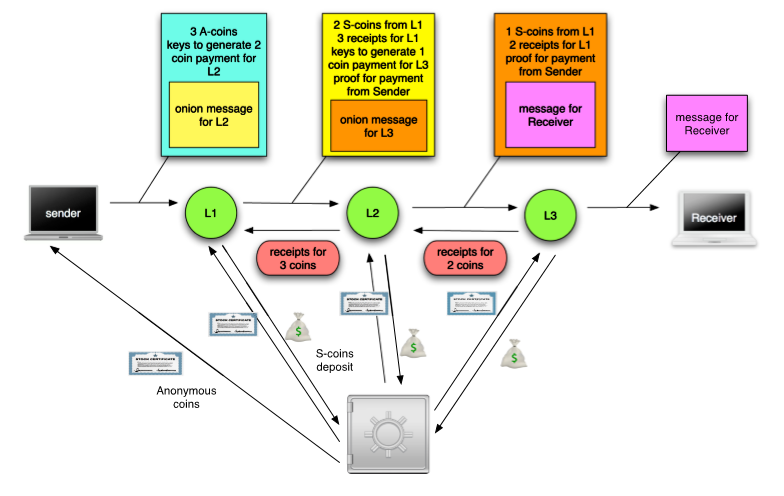}
  \caption{Hybrid payment scheme by Androulaki, E. (2008) \cite{PAR}}
\end{figure}

The system proposed by Ghosh,M, 2014 \cite{TorPathtoTorCoin} makes use of two systems: TorCoin and Torpath. The TorPath protocol assigns Tor circuits to clients, replacing the usual Tor directory servers with assignment servers which form decentralized consensus groups. In TorPath no client can generate its own circuit and no client can know the circuit of another client. The TorPath protocol has three stages: 1) Group Initialization: Assignment servers let clients connect to them and the assignment servers form consensus groups of the clients. Each relay chooses its position in the TOR circuit (Entry, Middle, Exit). For every position the client is asked to generate multiple public keys to potentially participate in multiple circuits on this position. 2) Verifiable Shuffle: The keys are shuffeled among different circuits to prevent the possibility to link keys to relays or clients. 3) Path Lookup: The client obtains the IP address of the entry relay and each relay obtains the IP address of its neighbouring relay in the circuit. The IP addresses are encrypted with the private key of each relay. Also, each circuit in the concensus group obtains a unique circuit identifier. The TorPath protocol ensures anonymity and circuit diversity. With TorCoin a protocol is proposed in which every relay in the tor circuits may mine a limit number of TorCoins as a reward for their contribution to the network. The mining of torcoins is initiated by the client: every $m$ messages send by the client the client starts the protocol to mine new torcoins. The clients have to be checked whether they indeed only mine new coins after $m$ messages. The authors say this is monitored by the assignment servers but no clear explanation is given of how this happens. If clients collude with each other this can give a security threat. When half of the tor relays are owned by an adversary 1/16 of assigned circuits will be owned by the adversary.

\begin{figure}
  \includegraphics[width=\linewidth]{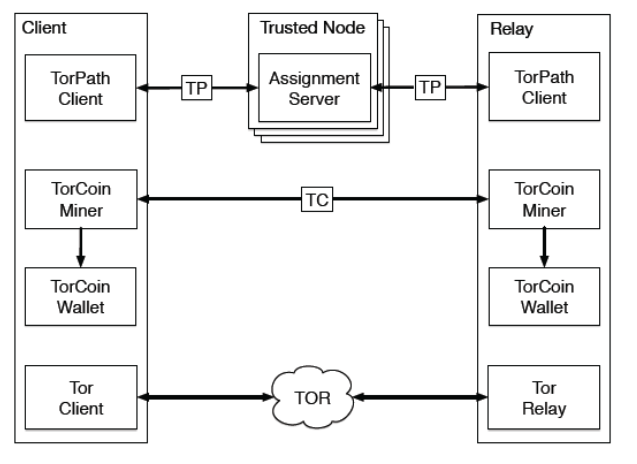}
  \caption{High level TorCoin and TorPath architecture \cite{TorPathtoTorCoin}}
\end{figure}

In another research by Dingledine, R. et al (2010) \cite{dingledine2010building} a solution is presented in which a 'gold star' is given to relays that provide good services for others. A gold star relay's traffic is given higher priority by other relays. The bandwidth is audited by the existing directory authorities to give users gold stars. After experimentation it is shown that nodes who are cooperative and thus share bandwidth and forward all gold-star traffic according to the rules have a faster download time and lower ping time. No practical implementation is given where it is tested whether users are indeed willing to contribute in exchange for a better service. 

Another such a system is the TEARS system proposed by Rob Jansen et al (2010) \cite{TEARS} and the BRAIDS system by Jansen, R. et al (2010) \cite{RecruitingTorRelays}. In the TEARS system are Band-with contributions rewarded with Shallots. Users can exchange Shallots for PriorityPasses to gain traffic priority. Shallots can be traded with other users. Open problems are with making incentives to participate, market economics policies, community effects and with deployment. Also the problem to determine if a relay was honest or not is not solved. BRAIDS introduces a ticket system which users can obtain from a bank and can be embedded into Tor cells to request services. The tickets are distributed by agent nodes that monitor other nodes. The agent nodes distribute tickets from the bank in proportion to the provided bandwidth. Each relay verifies it’s tickets to prevent double spending. A discrete event based simulator is used to show that their is an increased performance in traffic. With both TEARS and BRAIDS no implementation to test the system is given. 

A more complex solution to the free riding problem in TOR networks is the LIRA system proposed by Jansen, R. (2010) \cite{LIRA}. LIRA produces incentives with a novel cryptographic lottery design together with a new circuit scheduling algorithm that prioritizes traffic from those winning the lottery. Relays acquire electronic coins from the bank by providing service to the network. These coins can be exchanged for guaranteed winning tickets in the lottery and therefore provide in prioritized traffic in the TOR network. Other clients can also guess winning tickets with tune-able probability. Relays cannot distinguish from a guessed winner and a payed winner and thus maintain anonymity for paying clients. Mathematical arguments are given that LIRA provides economic incentives to buy tor usage, however no experiments are given in which LIRA is in use and there is a good working economy.

\subsection{Decentralized market design}
In 1996 Chavez and Maes propose the Kasbah agent based marketplace for buying and selling goods. A selling agent is anonymous, once it is put into the marketplace it negotiates on its own on behalf of the user. The user adds parameters to the agent like the desired date to sell an item, desired price and lowest acceptable price. The goal of the agent is to sell the item for the highest possible price. Once an agent has negotiated upon a deal the user can give final approval. At the same time their are buyers agents for which users also can set the following parameters: desired price, highest acceptable price and date to buy the item. The marketplace allows agents to interact with each other where all agents speak the same common language. The strategy of the buying and selling agents would be to first try to negotiate on the desired price. If this does not work, the agents will decrease or increase the price until a bargain is struck. Kasbah is a simple prototype to test the basic concepts. At the time of building Kasbah a truly useful system has yet to be made. \cite{Kasbah}

AVALANCHE is a prototype of an agent-based secure electronic commerce marketplace environment introduced by Padovan, B. et al in 2001. It is more advanced than Kasbah, it contains a reputation system. The authors envisage an Internet information ecosystem where billions of agent software systems interact with each other and humans to exchange a variety of information goods and services. The reputation of an agent is described as the amount of trust that is inspired by a particular person or agent inside the trading setting. By analyzing an agents previous cooperation behavior, the reputation and thus the trust of that agent is determined. In AVALANCHE agents communicate with a-symmetric public key encryption with each other to ensure confidentiality. However, the authentication of the agents is not guaranteed. It cannot be said whether an agent is a pseudonym or a real human. This problem is described as the Sybil-attack which is the most common attack to reputation systems. The strategies of the agents used differ from KASBAH in the sense that reputations of other agents are taken into consideration. Also at the end of a transaction the reputation of the other agent is changed accordingly to the success of the transaction. \cite{padovan2002prototype}

\begin{figure}
  \includegraphics[width=\linewidth]{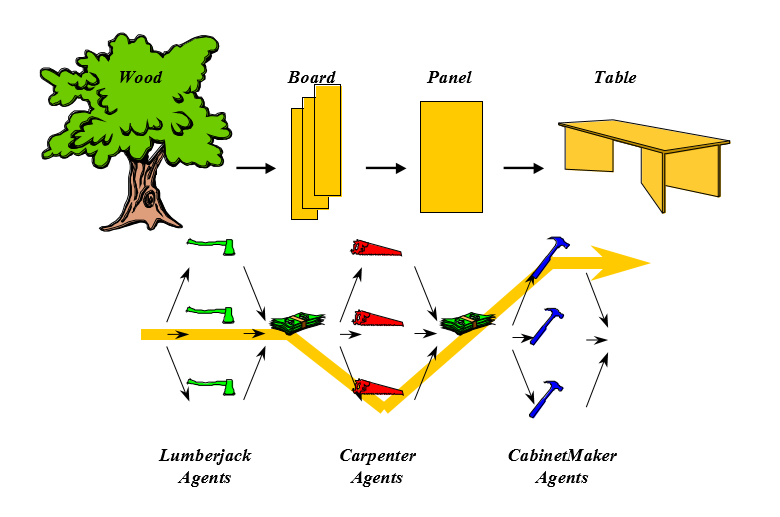}
  \caption{Avalanche value chain from tree to desk \cite{padovan2002prototype}}
\end{figure}

Soska, K. et al (2014) \cite{soska2016beaver} introduces a formal model for a decentralized anonymous marketplace (DAM), and the design of Beaver, a Sybil resistant DAM. The transactions and reviews of items in the marketplace are public, the relationship between the transaction and reviews are kept private and the customers in Beaver always remain anonymous. There are four basic transactions in Beaver 1) Registration: a vendor adds an item to the list of available items. 2) Payment: Funds are moved from a customer to a vendor. 3) Review: leave a review for an item. 4) Add transaction: add transcation to ledger. The ledger is a log of all the transactions which is maintained with bitcoin technology. The vendors first register themselves to the network, a customer can browse the different vendors and purchase an item from a vendor by doing an anonymous payment transaction. A customer can also give a review  by tighting a review to a payment transaction he made earlier. In the security thread model are two assumptions made: 1) 75\% of the nodes in the network need to be honest. 2) The customers and vendors are rational and do not behave maliciously if the cost of doing so is significant. Maybe say something about assumptions. For each transaction is a detailed algorithm described to perform the action. Fees are paid for each transaction and obtained by the node that adds the transaction to the ledger with bitcoin technology. Some limitations and points for future work are discussed like: vendor privacy(vendors might want to conceal their transaction volume) and values of fees. 

\section{Sybil attacks}
In a Sybil attack users create zero-cost identities to collude together and cheat the system. Lian, Q. et al, (2007) \cite{Collusion} measures user collusion activity in a real-world P2P system, Maze, that is fully controlled by the authors of the paper. Maze has a point system where peers consume points on downloading and gain points on uploading. A number of collusion detectors are identified: 1) Repetition-based collusion detection: Colluders generate large amounts of upload traffic with repeated content to generate points. The duplication degree is defined as the ratio of total upload traffic in bytes over the size of the unique data in bytes. When taking a look at the top 6 peers with the highest duplication degree it seems that the peers are colluding. The highest duplication degree of a peer is 43. The fact that the top 4 peers all generated duplicate traffic during the same days gives very strong evidence that the peers are colluding. 2) Group-based collusion detection: Groups of peers exchanging traffic to each other to generate points. It is possible for two friends to share large amounts of mutually interesting content. However, if these uploads to each other are relatively high comparing to their total traffic this is indicative for collusion. 3) Spam account collusion: One account is the main account which will get high points. Other pseudonym spam accounts are created with zero cost that will download from the main account to generate points for the main account.  4) Upload traffic concentration: If a lot of upload traffic from one peer is targeted towards a single physical machine this is a sign of collusion. The traffic concentration is measured:  the ratio of a peers highest upload traffic to a single machine to his total upload traffic. The Venn diagram shows that the collusion detectors sometimes mark the same peers as colluders. The detectors show empirical evidence that colluding behavior is happening in real world systems. It is yet not possible to provide definitive proof of the intent to collude. 

\begin{figure}
  \includegraphics[width=\linewidth]{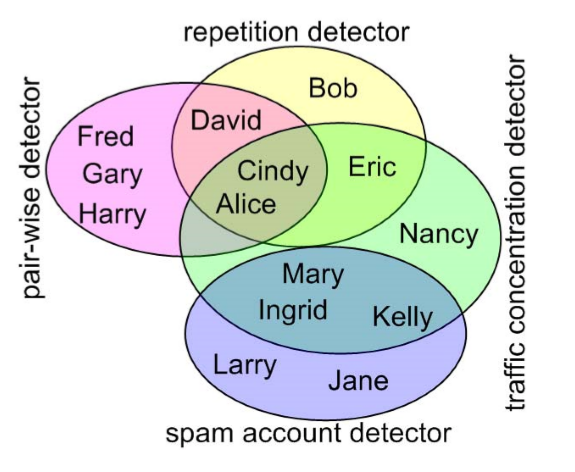}
  \caption{Comparing different collusion detectors\cite{Collusion}}
\end{figure}

In order to be protected against the Sybil attack in reputation systems for the trustworthiness of websites algorithms have been proposed by researchers. Hopcroft and Sheldon introduced the Global Hitting Time Mechanism (GHT) score in 2007 that is an improvement on the PageRank trust mechanism. \cite{hopcroft2007manipulation} In PageRank, the trustworthiness is based on multiple factors, but the most important factors are the number of links directing towards that website and the trustworthiness of a website that is referring. It is based on the principle that if more trustworthy people link toward a website, this website should be trustworthy. Pseudonyms can refer to each other to become trustworthy and these trustworthy pseudonyms can then increase the trust of certain websites by linking towards them. This is called the "two-loop attack". GHT differs from PageRank in that the links outgoing from the website of which the GHT score is determined are removed from the PageRank calculations. This protects against the two-loop attack. However GHT is still vulnerable to the restart-capture attack. The restart-capture attack make use of a vulnerability in the GHT algorithm when the calculation restarts at a different node. Thus Brandon, L. (2016) proposes a new algorithm called Personalized Hitting Time to solve this problem and improve the GHT algorithm. PHT works almost the same as GHT but calculates the score with a minor adjusted random walk. With an experiment is shown that PHT gives resistance against a specific kind of attack when agents show strategic behavior. Strategic behavior means that more sybils can be created by an agent. Also the informativeness of PHT remains high when more Sybils are added. But this property also remains high with Personalized PageRank. The definition of strategic behavior among agent is that a strategic agent creates misreports for other agents. Thus the agent will slander other agents with misreports. In PageRank this is equal to cutting outlinks to other pages or in other words to not create links to other pages. The strategic behavior is changed and redefined for every type of trust algorithm. In the PHT version no Sybils are added. The title of the paper: "Personalized Hitting Time for Informative Trust Mechanisms Despite Sybils" suggests that PHT provides an improvement on the trust mechanisms with Sybils. However, in the paper the strategic behavior of agents do not add any Sybils. Thus the impact of Sybils is not tested on PHT. \cite{liu2016personalized} \cite{levine2006survey} \cite{douceur2002sybil}

The lack of good algorithms to calculate trust scores with Sybils gave inspiration to Otte, P. to research Sybil-resistant trust mechanisms. Otte, P. introduces Temporal Page Rank, another random walk variant that makes use of a random jump in the random walk. An experiment in Tribler shows that a higher uploaded amount of data (trustworthiness) leads to a higher downloaded amount and thus that a fair trust mechanism is in place. However, there is no a strong guarantee against Sybil attacks. Another algorithm that is introduced by Otte, P. is the NetFlow algorithm that makes use of the Max-Flow algorithm of Ford-Folkerson to calculate a trust score. The informativeness is the percentage of agents that have a non zero reputation score. In NetFlow only agents with positive reputation scores will be granted resources. Some agents or new agents does not have uploaded enough resources in relation to their downloaded resources to have a positive reputation score which leads to low informativeness. To solve this a scaling of upload resources in comparison to download resources is implemented to generate a higher trust score among agents. This will allow weakly profitable Sybil attacks and is a trade-off with informativeness. \cite{SybilResTrust}

\section{Strategy-proofness in two-sided markets}
A "two sided market" is a market where two parties are linked together. For instance, a credit-card links consumers and merchants to each other  or newspapers link subscribers to advertisers. The software platforms that bring together these groups of users are considered a very important innovation and can be found in many industries.  Agents that operate in a two-sided market can develop strategies to exploit weaknesses in the market. \cite{eisenmann2006strategies} \cite{armstrong2006competition}

In 1962 Gale and Shapley introduced the first matching model in two sided markets. A preference list is a ranked list where the agent gives a preferred order of all the agent it wants to be matched with. For instance, a grain buyer provides an ordered ranked list of all the grain sellers it wants to buy from. With the deferred acceptance algorithm a "stable" match can be found. A "stable" match is a matching of all the buyers and sellers such that they can never do a better matching when re-matching each other later. The algorithm works as follows. Each buyer proposes to match itself to its preferred seller. A seller who receives multiple proposals from buyers chooses greedily the favourite buyer and rejects all other buyers. In the next stages each rejected buyer now proposes to their next choice and again sellers choose their most preferred option or reject otherwise. Gale and Shapley prove that this algorithm always lead to a stable matching. The deferred acceptance algorithm is a greedy algorithm because it makes the local optimal choice at each stage. \cite{gale1962college}

There are multiple ways in which matching can occur in two sided markets. In One-to-One matching there is one buyer matched to one seller like in the Gale and Shapley model of 1962. The matching mechanism is strategy-proof if truth-telling upon preference revelation in the deferred acceptance algorithm is a dominant strategy equilibrium when analyzing with game theory. According to Roth (1982) \cite{roth1982economics} there exists no matching algorithm that is both stable and strategy-proof for one-to-one matching problems. However, in the One-To-One case the proposing side (buyers) have truth-telling as a dominant strategy. Thus a stable matching is compatible with truth-telling for one side of the market. Another interesting thing is that Gale \& Sotomayor (1985) \cite{gale1985ms} showed that any stable matching in the One-to-One case is a Nash equilibrium in undominated strategies. An undominated strategy in game theory is a strategy where the outcome could be better or worse than another strategy depending on what other players do. This means that in the market case the side that receives matching proposals might be better off with another strategy than truth-telling depending on what the proposing side (buyers) do. \cite{abdulkadiroglu2013matching}

In Many-to-One matching multiple identities are matched to a single identity. For example, multiple students are matched to one college. Gale and Shapley already mentioned the college admissions type of problem and proposed an alternative deferred acceptance algorithm which proves to provide a stable matching. Also for this type of problem Roth showed in 1982 that there exists no mechanism that is stable and strategy-proof. \cite{roth1982economics} Roth also shows in 1986 that truth-telling is a weakly dominant strategy for all students under the student-optimal stable mechanism. For colleges that is different. There exists no stable mechanism where truth-telling is a weakly dominant strategy for all colleges. \cite{roth1986allocation} The student to college matching mechanism was used by the Boston School Committee because the original system would incentivize to "game the system". It enables families to list their true choices of schools without jeopardizing their chances of being assigned to any school by doing so. The Boston School example shows that the two sided market theory is successfully implemented in the real world.\cite{abdulkadiroglu2013matching} \cite{rysman2009economics}

\section{Conclusions}
Building a decentralized market requires a trust system that tell peers the trustworthiness of other peers. Building these trust systems proofs to be a challenging problem because there are various solutions both in academic designs and real world implementations each with their own problems. A wide used solution is building a reputation system that gives users an indication of the trustworthiness of other users. Most reputation systems in both the academic literature and real world implementations meet their trust related design requirement but are not defended against the Sybil attack. In the Sybil attack fake identities are created by an adversary to manipulate the trust scores of users. It is shown that it is impossible to completely defend reputation systems against the Sybil attack, but there is ongoing research to make reputation systems more resilient against the Sybil attack. Another proposed solution for a trust mechanism is the use of smart contracts where people can transfer money via a trustless public ledger with programmable contract details. Smart contracts should eventually be able to replace paper contracts. Another challenge in decentralized markets is the problem of users being able to use strategies to manipulate the market. This can prevented with a good matching engine.
\clearpage
\bibliographystyle{abbrv}
\bibliography{sample}
%\bibliography{biblio}

\end{document}